\documentclass[aps,pra,twocolumn,superscriptaddress,showpacs]{revtex4}
\usepackage{bm}
\usepackage{epsf}
\usepackage{amssymb}
\usepackage{graphicx}
\usepackage{rotating}
\usepackage{epsfig}
\usepackage{psfrag}
\begin{document}

\def\ba{\mathbf{a}}
\def\d{\mathbf{d}}
\def\P{\mathbf{P}}
\def\bK{\mathbf{K}}
\def\bk{\mathbf{k}}
\def\bkn{\mathbf{k}_{0}}
\def\bx{\mathbf{x}}
\def\bfn{\mathbf{f}}
\def\bg{\mathbf{g}}
\def\bj{\mathbf{j}}
\def\bR{\mathbf{R}}
\def\br{\mathbf{r}}
\def\bu{\mathbf{u}}
\def\bq{\mathbf{q}}
\def\bw{\mathbf{w}}
\def\bp{\mathbf{p}}
\def\bG{\mathbf{G}}
\def\bz{\mathbf{z}}
\def\bs{\mathbf{s}}
\def\E{\mathbf{E}}
\def\bv{\mathbf{v}}
\def\b0{\mathbf{0}}
\def\la{\langle}
\def\ra{\rangle}
\def\beq{\begin{equation}}
\def\eeq{\end{equation}}
\def\bea{\begin{eqnarray}}
\def\eea{\end{eqnarray}}
\def\bdm{\begin{displaymath}}
\def\edm{\end{displaymath}}
\def\bnab{\bm{\nabla}}

\title{Variational theory of two-fluid hydrodynamic modes at unitarity}

\author{E.~Taylor}
\affiliation{Department of Physics, University of Toronto, Toronto, Ontario,
Canada, M5S 1A7}
\author{H.~Hu}
\affiliation{Department of Physics, Renmin University of China, Beijing 100872, China}
\affiliation{ARC Centre of Excellence for Quantum-Atom Optics, Department of Physics, University of Queensland, Brisbane, Queensland 4072, Australia}
\author{X.-J.~Liu}
\affiliation{ARC Centre of Excellence for Quantum-Atom Optics, Department of Physics, University of Queensland, Brisbane, Queensland 4072, Australia}
\author{A.~Griffin}
\affiliation{Department of Physics, University of Toronto, Toronto, Ontario,
Canada, M5S 1A7}

\date{\today}

\begin{abstract}
We present the results of a variational calculation of the frequencies of the low-lying Landau two-fluid hydrodynamic modes in a trapped Fermi superfluid gas at unitarity.  Landau's two-fluid hydrodynamics is expected to be the correct theory of Fermi superfluids at finite temperatures close to unitarity, where strong interactions give rise to collisional hydrodynamics.  Two-fluid hydrodynamics predicts the existence of in-phase modes in which the superfluid and normal fluid components oscillate together, as well as out-of-phase modes where the two components move against each other.  We prove that at unitarity, the dipole and breathing in-phase modes are locally isentropic.  Their frequencies are independent of temperature and are the same above and below the superfluid transition.  The out-of-phase modes, in contrast, are strongly dependent on temperature and hence, can be used to test the thermodynamic properties and superfluid density of a Fermi gas at unitarity.  We give numerical results for the frequencies of these new modes as function of temperature in an isotropic trap at unitarity.
\end{abstract}


\maketitle

\section{Introduction}
\label{intro}
Landau's two-fluid hydrodynamics~\cite{Landau41, Khalatnikov} is the theory of the finite temperature dynamics of all superfluids (with a two-component order parameter) when collisions are sufficiently strong to produce a state of local thermodynamic equilibrium.   Recent experiments have begun to probe the collective modes in trapped superfluid Fermi gases with a Feshbach resonance~\cite{Thomas04,Grimm04}.  At unitarity, the magnitude of the $s$-wave scattering length $a_s$ that characterizes the interactions between fermions in different hyperfine states diverges ($|a_s|\rightarrow \infty$).  Owing to the strong interaction close to unitarity, we expect that the dynamics of superfluid Fermi gases with a Feshbach resonance at finite temperatures are described by Landau's two-fluid hydrodynamic equations~\cite{TaylorPRA05}.  

Solving Landau's two-fluid equations for trapped gases is difficult due to the fact that the density profiles of the superfluid and normal fluid components are highly nonuniform, making a reliable ``brute-force" numerical calculation very challenging~\cite{Ho98,Levin07}.  In a recent paper~\cite{TaylorPRA05}, an alternate variational formulation of Landau's two-fluid equations was developed.  Following the approach pioneered by Zaremba~\textit{et al.}~\cite{ZNG}, we use a simple ansatz for the superfluid and normal fluid velocity fields based on exact solutions at $T=0$ and above $T_c$.  This gives algebraic equations for the variational parameters describing the breathing and dipole two-fluid modes.  The coefficients in these equations involve spatial integrals over equilibrium thermodynamic quantities.  This approach is simpler than solving the two-fluid equations directly for trapped gases.  In the present paper, we report numerical results for the breathing and dipole mode frequencies at unitarity for an isotropic trap based on this variational method.  However, our general approach can also be used away from unitarity.

We discuss the in-phase breathing mode at unitarity since this mode has been studied extensively in recent experiments~\cite{Thomas04,Grimm04,Thomas05}.  In particular, we examine the surprising results of the experiments by Thomas and coworkers~\cite{Thomas05} that have shown the frequency of this in-phase mode to be almost independent of temperature, remaining within a few percent of its $T=0$ value even well above the superfluid transition temperature $T_c$.  Our analysis of the Landau two-fluid equations at unitarity shows that the in-phase breathing and dipole hydrodynamic modes are locally isentropic, mode, with the superfluid and normal fluid moving with the same velocity, $\bv_s(\br,t) = \bv_n(\br,t)$.  We find that the frequencies of these in-phase modes are independent of temperature, given by their $T=0$ value at all temperatures.  

Of greater interest are the \textit{out-of-phase} breathing and dipole modes, which have not been studied experimentally.  These modes involve an oscillation of the trapped superfluid where the superfluid and normal fluid components move against each other, in contrast to the in-phase modes where these components move together.  The out-of-phase modes are predicted to be strongly temperature-dependent and should provide a useful tool to test the microscopic model used for the thermodynamic properties.

In a companion paper~\cite{HuLiuPRL08}, we show how these two-fluid modes can be measured using standard two-photon Bragg scattering techniques~\cite{DavidsonRMP05}.  Extending the variational method described in this paper, we show the density response function has resonances at the breathing and dipole mode frequencies.  

In our variational theory~\cite{TaylorPRA05}, calculation of the frequencies of the two-fluid modes requires knowing the values of a number of thermodynamic quantities.  At unitarity, however, the variational equations simplify with only two thermodynamic quantities required for the dipole and breathing mode frequencies: the superfluid density $\rho_s$ and the isentropic compressibility $(\partial \mu/\partial\rho)_{s}$.  In this paper, we calculate the latter quantity at unitarity using the fluctuation theory developed in Ref.~\cite{HLD}, which is an improved version of the original theory of Nozi\`eres and Schmitt-Rink (NSR)~\cite{NSR}.  As shown in Ref.~\cite{HLD}, this theory gives thermodynamic quantities at finite temperatures which are in excellent agreement with \textit{ab-initio} calculations~\cite{Giorgini,Bulgac,Svistunov} and recent experimental measurements~\cite{HDL2007}.  The superfluid density we use is also based on the NSR fluctuation theory~\cite{TaylorPRA06,TaylorPRA07}.  The spatially-varying compressibility and superfluid density that enter our variational two-fluid equations are calculated within a local density approximation (LDA) using our results for a uniform Fermi superfluid.

He {\textit{et al.}}~\cite{Levin07} have also reported results for the two-fluid modes in an isotropic trap, based on a direct numerical solution of the Landau two-fluid differential equations. While there is some ambiguity in identifying the nature of the oscillations in Ref.~\cite{Levin07}, the in-phase breathing mode is found to be temperature-independent, in agreement with our variational results.  However, the temperature dependence of the out-of-phase mode breathing mode is very different from what we obtain (see Section~\ref{bmu}).

Heiselberg~\cite{Heiselberg05} has discussed the first and second sound velocity in the BCS-BEC crossover for a uniform gas.  In this case, the solutions of the two-fluid equations are known (plane waves).  For the thermodynamic functions which are needed, Heiselberg worked these out in the BCS and BEC limits and interpolated these results to describe the unitarity region.  Our work makes a major extension of this previous study since we deal with a non-uniform trapped superfluid and use a microscopic theory for the thermodynamic functions and the superfluid density at unitarity.  

In Section~\ref{thermunisec}, we discuss some of the features of ``universal" thermodynamics valid at unitarity~\cite{Ho04}.  We use these results in Section~\ref{isentropic} to prove that Landau's two-fluid hydrodynamic equations predict a locally isentropic breathing mode at unitarity, corresponding to a situation where both the normal and superfluid components move with the same local velocity.  In Section~\ref{review}, we review the variational formulation of Landau's two-fluid equations given in Ref.~\cite{TaylorPRA05}. In Section~\ref{NSRsec}, we discuss the NSR results for the temperature dependent isentropic compressibility and superfluid density which we need as inputs in our variational solutions. In Section~\ref{bmf}, we reformulate the equations for the breathing modes derived in Ref.~\cite{TaylorPRA05} in a more useful form for use at unitarity.   In Section~\ref{bmu}, we show that the predictions of universal thermodynamics allow us to derive simple expressions for the breathing mode frequencies at unitarity.   Numerical results for the temperature dependence of the frequency of the out-of-phase \textit{breathing} mode are also given for a trapped gas using a local density approximation (LDA).  In Section~\ref{secdipole}, we calculate the temperature dependence of the out-of-phase \textit{dipole} mode frequency.

In Appendix~\ref{appendixA}, we compare the isentropic breathing mode in trapped Fermi superfluid gases with first-sound in superfluid $^4$He, which is also a locally isentropic mode. 
Appendices~\ref{appendixD} and~\ref{appendixC} discuss the low and high temperature limits of the frequency of the out-of-phase breathing mode using a BCS mean-field theory (without fluctuations). These calculations confirm the main features of the LDA results given in the text, still within the same variational ansatz.

\section{Thermodynamics at unitarity}
\label{thermunisec}

In this Section, we review the features of universal thermodynamics at unitarity~\cite{Ho04} and use these to derive a number of thermodynamic identities at unitarity that will be used throughout this paper.  

In a dilute, uniform system of interacting fermions, there are three microscopic length scales (for a recent review and references on Fermi gases, Giorgini, Pitaevskii, and Stringari~\cite{Trentoreview}).  The three length scales are the mean interparticle spacing $n^{-1/3}_F$, the thermal wavelength $\lambda^2_T \equiv 2\pi/mk_BT$ (throughout this paper we set $\hbar =1$), and the $s$-wave scattering length $a_s$ that completely characterizes the interaction between different species (denoted by the $\uparrow,\downarrow$) of fermions in the low-density limit.  Here, $n_F \equiv (2m\epsilon_F)^{3/2}/3\pi^2$ is the density of both species of fermions (i.e., $n_F = n_{\uparrow}+n_{\downarrow}$), where $\epsilon_F$ is the Fermi energy of an ideal gas.  The corresponding energy scales are the kinetic energy $\epsilon_F$, $k_BT$, and the interaction energy (which can be expressed as a functional of the density $n_F$ and $a_s$).  At unitarity, the scattering length diverges, meaning that the only remaining length scales are the interparticle spacing $n^{-1/3}_F$ and the thermal wavelength, as first argued by Ho~\cite{Ho04}.  This also implies that at unitarity, the only energy scales are the Fermi energy and $k_BT$.  Consequently, the only dimensionless energy scale at unitarity is $k_BT/\epsilon_{F} \equiv k_BT/k_BT_F$.  This immediately means that all thermodynamic functions at unitarity can be written in dimensionless form as a function of the ratio $T/T_F$.  These features can be used to derive useful identities involving the internal energy, entropy, and chemical potential.

Owing to the fact that there is only one dimensionless energy scale, given by $k_BT/k_BT_F(\rho)$, the internal energy density $U$ in a trapped Fermi gas takes the form~\cite{Ho04,Thomas05}
\bea U = \frac{\rho\epsilon_F(\rho)}{m}f_E[T/T_F(\rho)].\label{Uuni}\eea Also, the total entropy $S$ of a fluid element of small (infinitesimal) volume $\Delta V$ is~\cite{Ho04,Thomas05}
\bea S = N k_B f_S[T/T_{F}(\rho)].\label{S}\eea Here $f_E$ and $f_S$ are dimensionless functions of the reduced temperature $T/T_F(\rho)$.  $\epsilon_F(\rho)$ is the local Fermi energy and is a function of the mass density $\rho(\br)$.  $N(\br) = \rho(\br) \Delta V/m$ is the total number of fermions in the small volume $\Delta V$ centered at position $\br$.   We emphasize that both the energy density $U(\br)$ and the entropy $S(\br)$ of a small fluid volume centered at $\br$ depend on position through the Fermi energy $\epsilon_F(\rho)$ and the local mass density $\rho(\br)$.  

The total local energy density is given by $E_0 = U + \rho V_{\mathrm{ext}}$, where
\bea V_{\mathrm{ext}} = \frac{1}{2}\sum_i\omega^2_ix^2_i \label{trap}\eea
is the harmonic trapping potential divided by the mass.  It is standard for Landau's two fluid equations to be given in terms of the mass density $\rho=mn$, instead of the number density $n$.  Thus we use this scaled harmonic trap potential.  At unitarity, the energy of the small volume $\Delta V$ is thus
\bea E_0\Delta V = N\epsilon_F(\rho)f_E[T/T_F(\rho)] + NmV_{\mathrm{ext}}.\eea
The pressure $P$ is defined by
\bea P = -\left(\frac{(E_0 \Delta V)}{\partial \Delta V}\right)_{\!\!N,S}. \eea
From Eq.~(\ref{S}), we see that holding $N$ and $S$ constant requires holding the reduced temperature constant as well~\cite{Thomas05}.  Thus we find
\bea
\left(\frac{(E_0 \Delta V)}{\partial \Delta V}\right)_{\!\!N,S} &=& N\left(\frac{\partial\epsilon_F(\rho)}{\partial \Delta V}\right)_{\!\!N}f_E[T/T_F(\rho)]\nonumber\\&=&-\frac{\rho^2}{m}\frac{\partial \epsilon_F(\rho)}{\partial \rho}f_E[T/T_F(\rho)]
\nonumber\\&=&-\frac{2}{3}\frac{\rho\epsilon_F(\rho)}{m}f_E[T/T_F(\rho)].\eea
Thus, at unitarity, the pressure and energy density are related by~\cite{Thomas05}
\bea P = \frac{2}{3}\frac{\rho\epsilon_F(\rho)}{m}f_E[T/T_F(\rho)] = \frac{2}{3}U,\label{P23U}\eea 
the same relation one obtains in a noninteracting Fermi or Bose gas.  The temperature is defined by \bea 
T = \left(\frac{\partial U}{\partial s}\right)_{\!\rho},\label{Tdef}\eea where $s = S/\Delta V$ is the entropy density.  Using this, Eq.~(\ref{P23U}) implies that
\bea \frac{\partial}{\partial x_i}\left(\frac{\partial P}{\partial s}\right)_{\!\!\rho} = \frac{2}{3}\frac{\partial T_0}{\partial x_i} = 0,\label{identity4}\eea since the equilibrium temperature $T_0$ is spatially uniform, even in a harmonically confined gas with nonuniform density.  

The chemical potential per unit mass is given by~\cite{TaylorPRA05} 
\bea \mu = \left(\frac{\partial U}{\partial\rho}\right)_{\!s}\!\! + \;V_{\mathrm{ext}}. \label{mudef0}\eea 
Combining this expression with Eq.~(\ref{P23U}), we also obtain
\bea \left(\frac{\partial P}{\partial \rho}\right)_{\!s} = \frac{2}{3}\left[\mu - V_{\mathrm{ext}}\right].\label{Punitid}\eea
Using this, we find \bea \frac{\partial}{\partial x_i}\left(\frac{\partial P}{\partial \rho}\right)_{\!s} = \frac{2}{3}\frac{\partial \mu_{0}}{\partial x_i} - \frac{2}{3}\frac{\partial V_{\mathrm{ext}}}{\partial x_i} = -\frac{2}{3}\omega^2_ix_i.\label{identity5}\eea
Here we have made use of that fact that, like the temperature $T_0$, the equilibrium chemical potential $\mu_0$ is spatially uniform, $\bnab\mu_{0} = 0$.

We will make use of the identities derived in this Section (for a Fermi gas at unitarity) throughout this paper.

\section{Locally isentropic dynamics}
\label{isentropic}
Before discussing our variational solutions of the two-fluid equations in Section~\ref{review}, we use the results of Section~\ref{thermunisec} to discuss some general features of the solutions of the Landau two-fluid hydrodynamic equations for trapped superfluid gases.  
In particular, Thomas \textit{et al.}~\cite{Thomas05} argued that the (in-phase) breathing mode at unitarity obeys a single Euler equation for the velocity $\bv \equiv \bv_s = \bv_n$ on the grounds of \textit{locally isentropic hydrodynamics}.  It followed from the analysis of this Euler equation that the frequency of the breathing mode would be independent of temperature.  This surprising result was consistent with their experimental results for the breathing mode.  We now derive this starting from Landau's two-fluid hydrodynamic equations.  

We start with the continuity and conservation of entropy equations of Landau two-fluid hydrodynamics~\cite{Khalatnikov},
\bea \frac{\partial \rho}{\partial t} + \bnab\cdot\bj=0\label{rhocont}\eea and 
\bea \frac{\partial s}{\partial t} + \bnab\cdot\left(s\bv_n\right) = 0.\label{scont}\eea  The total mass current \bea \bj = \rho_s\bv_s + \rho_n\bv_n\label{bj}\eea is given in terms of the superfluid and normal fluid velocities $\bv_s$ and $\bv_n$, as well as the superfluid and normal fluid densities, $\rho_s$ and $\rho_n$.  The sum of the superfluid and normal fluid densities gives the total mass density, $\rho = \rho_s + \rho_n$. 
The continuity equation in Eq.~(\ref{rhocont}) expresses mass conservation and is always valid.  Equation~(\ref{scont}) assumes that the entropy of the fluid is carried by the normal fluid and is conserved.  These equations describe reversible flow without any dissipation arising from transport coefficients~\cite{Khalatnikov}.  

An oscillation is \textit{locally} isentropic if the entropy per unit mass $\bar{s}(\br,t) \equiv s(\br,t)/\rho(\br,t) = S(\br,t)/\rho(\br,t)\Delta V$ does not change in time as the mass element $\rho(\br,t)\Delta V$  moves with the fluid.  Defining the Lagrangian derivative
\bea \frac{\mathrm{D}}{\mathrm{D} t} \equiv \frac{\partial}{\partial t} + \bv\cdot\bnab,\label{Lder}\eea
locally isentropic hydrodynamics corresponds to the situation where
\bea \frac{\mathrm{D}\bar{s}}{\mathrm{D} t} = 0.\eea
Using Eqs.~(\ref{rhocont}) and (\ref{scont}), one can show that
\bea \frac{\partial \bar{s}}{\partial t} + \bv_n\cdot\bnab\bar{s}= \frac{\bar{s}}{\rho}\bnab\cdot\rho_s(\bv_s - \bv_n).\label{barsdt}\eea
This result confirms that the dynamics of a fluid are locally isentropic when $\bv_s = \bv_n\equiv \bv$.  

For locally isentropic fluid flow, Landau's expression for the current in Eq.~(\ref{bj}) reduces to $\bj = (\rho_s + \rho_n)\bv = \rho\bv$.  Using this result in Landau's equation of motion for the current~\cite{TaylorPRA05}
\bea
\frac{\partial \bj}{\partial t} &=& -\bnab P - \rho\bnab V_{\text{ext}} -
\rho_s \bv_s\cdot \bnab \bv_s - \rho_n\bv_n\cdot\bnab\bv_n\nonumber\\
&&-\bv_s \bnab\cdot (\rho_s \bv_s) - \bv_n \bnab\cdot (\rho_n \bv_n), \label{jt} \eea
it reduces to
\bea
\frac{\partial \bj}{\partial t} &=& -\bnab P - \rho\bnab V_{\text{ext}} -
\rho \bv\cdot \bnab \bv-\bv \bnab\cdot \bj. \label{jt2}\eea
Combining this equation with the continuity equation given by Eq.~(\ref{rhocont}), we obtain the following equation of motion for the velocity $\bv$:
\bea \frac{\partial \bv}{\partial t} = -\bnab\left(\frac{\bv^2}{2} + V_{\mathrm{ext}}\right) - \frac{\bnab P}{\rho}.\label{Euler}\eea
This is precisely Euler's equation for an ideal \textit{irrotational} (such that $\bnab \bv^2 = 2\bv\cdot\bnab\bv$) fluid~\cite{LLFM}, generalized to include the effects of an external trapping potential.  
This result shows that for the special case where $\bv_s(\br,t) = \bv_n(\br,t)$, Landau's two-fluid hydrodynamic equations reduce to Euler's equation for an irrotational velocity field.  

Our present discussion shows the equation of motion considered in Ref.~\cite{Thomas05} is a rigorous consequence of Landau's two-fluid equations for locally isentropic flow.  We now derive a condition for a locally isentropic ($\bv_s = \bv_n$) normal mode solution of the Landau two-fluid equations to exist.  

The linearized continuity and entropy conservation equations [given by Eqs.~(\ref{rhocont}) and (\ref{scont})] are
\bea \frac{\partial \delta \rho}{\partial t} +
\bnab\cdot\left(\rho_{s0}\bv_s +
\rho_{n0}\bv_n\right)=0 \label{rhocontl}\eea
and
\bea \frac{\partial \delta s}{\partial t} 
+\bnab\cdot\left(s_0\bv_n\right)=0.\label{scontl}\eea
Introducing the displacement fields~\cite{TaylorPRA05,ZNG} $\bu_s,\bu_n$,
\bea
\bv_s(\br,t) \equiv \frac{\partial \bu_s(\br,t)}{\partial t}, \; \; \;
\bv_n(\br,t) \equiv \frac{\partial \bu_n(\br,t)}{\partial t}, \label{fields}
\eea the linearized continuity and entropy conservation equations can be
expressed in terms of these 
fields as\bea \delta
\rho(\br,t) = -\bnab\cdot\left[\rho_{s0}(\br)\bu_s(\br,t) +
\rho_{n0}(\br)\bu_n(\br,t)\right] \label{rhocontdf} \eea and
\bea\delta s(\br, t) &=&
-\bnab \cdot \left[s_0(\br)\bu_n(\br,t)\right].\label{scontdf} \eea 
These expressions will be used in deriving the conditions for a locally isentropic mode to exist.  

Since each mass element evolves at constant entropy in a locally isentropic flow, these elements do not exchange heat with their surroundings and hence the temperature remains unchanged throughout the fluid.  
From the linearized Landau two-fluid equations for the superfluid and normal fluid densities (see Eqs.~(38) and (39) in Ref.~\cite{TaylorPRA05}), one can show that
\bea \frac{\partial(\bv_s-\bv_n)}{\partial t} = \frac{s_0}{\rho_{n0}}\bnab\delta T.\eea
This implies $\bnab\delta T=0$ when $\bv_s = \bv_n$, showing that the temperature remains constant everywhere for a locally isentropic mode.  Thus, a locally isentropic mode is also a locally \textit{isothermal} mode.  Using $\delta T = (\partial T/\partial s)_{\rho}\delta s + (\partial T/\partial \rho)_{s}\delta\rho$ and Eqs.~(\ref{rhocontdf}) and (\ref{scontdf}), we can write the condition $\bnab\delta T = 0$ as
\bea \bnab\left[\left(\frac{\partial T}{\partial\rho}\right)_{\!s}\bnab\cdot(\rho_0\bu) + \left(\frac{\partial T}{\partial s}\right)_{\!\rho}\bnab\cdot(s_0\bu)\right] = 0,\label{condition0}\eea
where $\bu_s = \bu_n \equiv \bu$.  

To make contact with the results of Section~\ref{thermunisec}, we express Eq.~(\ref{condition0}) in terms of derivatives of the pressure.  The pressure can be expressed in terms of the equilibrium thermodynamic identity~\cite{TaylorPRA05},
\bea P = -U -
\rho V_{\mathrm{ext}} + Ts + \mu\rho. \eea
Treating $P,T$, and $\mu$ as functions of the independent variables $\rho$ and $s$, using the Maxwell relation
\bea \left(\frac{\partial T}{\partial\rho}\right)_{\!s} = \left(\frac{\partial \mu}{\partial s}\right)_{\!\rho},\label{maxwell}\eea  and Eqs.~(\ref{Tdef}) and (\ref{mudef0}), one can show that
\bea \left(\frac{\partial P}{\partial \rho}\right)_{\!s}&=&\rho_0\left(\frac{\partial \mu}{\partial\rho}\right)_{s} + s_0\left(\frac{\partial \mu}{\partial s}\right)_{\rho} \label{identity2}\eea and
\bea \left(\frac{\partial P}{\partial s}\right)_{\!\rho}&=&\rho_0\left(\frac{\partial T}{\partial\rho}\right)_{s} + s_0\left(\frac{\partial T}{\partial s}\right)_{\rho}. \label{identity3}\eea
The gradient of the equilibrium temperature $T_0$ can be written as
\bea \bnab T_0 &=&\left(\frac{\partial T}{\partial \rho}\right)_{\!s}\bnab\rho_0 + 
\left(\frac{\partial T}{\partial s}\right)_{\!\rho}\bnab s_0=0 .
\label{bnabT0}\eea
This gives the following useful identity for a trapped gas:
\bea\left(\frac{\partial
T}{\partial \rho}\right)_{\!s} 
\frac{\partial \rho_0}{\partial x_j} + \left(\frac{\partial
T}{\partial s}\right)_{\!\rho}\frac{\partial s_0}{\partial x_j}=0. \label{dTdx}\eea

Using Eqs.~(\ref{identity3}) and (\ref{dTdx}), the condition in Eq.~(\ref{condition0}) can be rewritten in the useful form
\bea \bnab\left(\bnab\cdot\bu\right)\left(\frac{\partial P}{\partial s}\right)_{\!\rho}+ \left(\bnab\cdot\bu\right)\bnab\left(\frac{\partial P}{\partial s}\right)_{\!\rho}=0.\label{decouplecondition}\eea  
Equation~(\ref{decouplecondition}) thus gives the condition for there to exist a locally isentropic (or isothermal) normal mode solution of the Landau two-fluid equations.  This relation is completely general for an oscillation described by $\bu$.  It is not restricted to the case of a superfluid in a harmonic trap at unitarity, although this is the region of interest in this paper.  

At unitarity, the second term in Eq.~(\ref{decouplecondition}) vanishes in accordance with Eq.~(\ref{identity4}).  Thus we conclude that a locally isentropic mode ($\bv_s = \bv_n$) exists at unitarity if either
\bea \left(\frac{\partial P}{\partial s}\right)_{\!\rho}=\frac{2}{3}T=0, \label{decouplecondition2}\eea
or if
\bea \bnab(\bnab\cdot\bu) = 0 \label{decouplecondition3}\eea is satisfied.   
The first condition given by Eq.~(\ref{decouplecondition2}) is trivially satisfied at $T=0$.  Here the normal fluid vanishes and hence all particles move with the same velocity $\bv_s = \bv$, and of course any oscillation will be locally isentropic.  In order for a locally isentropic mode to exist at \textit{finite} temperatures, Eq.~(\ref{decouplecondition3}) must be satisfied.  This is satisfied by the scaling solution~\cite{scaling} $\bv(\br,t) \propto \br\cos\omega t$ [equivalently $\bu(\br,t) \propto \br\cos\omega t$] of the hydrodynamic equation in Eq.~(\ref{Euler}) that describes the breathing mode.  It is also satisfied by the \textit{generalized Kohn mode} (the in-phase dipole mode) that we discuss in Section~\ref{secdipole}.  This suggests that the existence of a purely locally isentropic mode is \textit{not} a universal feature of hydrodynamics at unitarity, but rather is a special feature in a harmonically confined gas.

In Section~\ref{bmu}, we confirm that our variational solution of the two-fluid equations gives a locally isentropic breathing mode with a frequency independent of temperature.  We call this breathing mode the ``in-phase" breathing mode since the normal and superfluid components move together, $\bv_s = \bv_n$.  This is the mode studied by Thomas and coworkers~\cite{Thomas05}.  In addition, our variational solution also predicts an out-of-phase breathing mode which is not locally isentropic, with a frequency very strongly dependent on temperature.  

In superfluid $^4$He, \textit{first sound} also describes a locally isentropic mode, a fact accounted for by Eq.~(\ref{decouplecondition}).  However, first sound in uniform superfluid $^4$He is locally isentropic for different reasons than the in-phase breathing and dipole modes in a trapped Fermi superfluid at unitarity.  This is discussed in Appendix~\ref{appendixA}.

\section{Variational solution of the two-fluid equations}
\label{review}

While the preceding analysis showed that the Landau two-fluid equations at finite temperatures admit a class of analytic solutions at unitarity [corresponding to $ \bnab(\bnab\cdot\bu)=0$], these solutions only describe the in-phase [$\bu_s = \bu_n \equiv \bu$] dipole and breathing mode oscillations.  The out-of-phase solutions of the two-fluid equations cannot be obtained using such a simple analysis.  We shall use a variational method to derive expressions for the frequencies of these out-of-phase modes.   
In this section, we review the variational formulation of Landau's two-fluid equations developed in Ref.~\cite{TaylorPRA05}.  

In 1950, Zilsel~\cite{Zilsel50} introduced a  phenomenological action $S[s,\rho,\rho_n,\bv_s,\bv_n]$ as a function of the entropy density $s$, the total density $\rho = \rho_n + \rho_s$, the normal fluid density $\rho_n$, as well as the superfluid $\bv_s$ and normal fluid $\bv_n$ velocities. By construction, the variation of this action with respect to these variables generates the Landau two-fluid equations.  In order to generate the \textit{linearized} two-fluid equations (the solutions of which determine the spectrum of normal modes), the action is expanded in powers of fluctuations ($\delta\rho,\delta s,\delta\bv_s,\delta\bv_n$) about the equilibrium values ($\rho_{s0},s_0,\bv_{s0},\bv_{n0}$) up to quadratic order.  We assume that $\bv_{s0} = 0$ and $\bv_{n0} = 0$, so that $\delta\bv_n = \bv_n$ and $\delta\bv_s = \bv_s$.  The terms in the action that describe fluctuations $\delta\rho_n$ in the normal fluid density can be shown to be higher-order~\cite{TaylorPRA05} and are thus neglected.  The resulting action describes the hydrodynamic fluctuations. It is further simplified by replacing the entropy and density fluctuations $\delta s$ and $\delta\rho$ in terms of the superfluid and normal fluid velocities.  This can be done using the linearized continuity and entropy conservation equations in Eqs.~(\ref{rhocontl}) and (\ref{scontl}).

Using Eqs.~(\ref{fields}), (\ref{rhocontdf}), and (\ref{scontdf}), the action that describes hydrodynamic fluctuations ($\delta\rho,\delta s,\bv_s,\bv_n$) can be expressed in terms of the two displacement fields $\bu_s$ and $\bu_n$~\cite{TaylorPRA05}, 
\bea
S^{(2)} &=& \int d\br dt\;\Bigg\{\frac{1}{2}\rho_{s0}\dot{\bu}_s^2 +
\frac{1}{2}\rho_{n0}\dot{\bu}_n^2 \nonumber\\&&-
\frac{1}{2}\left(\frac{\partial
\mu}{\partial\rho}\right)_{\!s}\left[\bnab\cdot\left(\rho_{s0}\bu_s
+\rho_{n0}\bu_n\right)\right]^2
\nonumber\\&&-\left(\frac{\partial T}{\partial
\rho}\right)_{\!s}\left[\bnab\cdot\left(s_0\bu_n\right)\right]\left[\bnab
\cdot\left(\rho_{s0}\bu_s
+\rho_{n0}\bu_n\right)\right]
\nonumber\\&&-
\frac{1}{2}\left(\frac{\partial T}{\partial
s}\right)_{\!\rho}\left[\bnab\cdot\left(s_0\bu_n\right)\right]^2\Bigg\}.
\label{S0c3} \eea 
Here $\mu$ is the chemical potential per unit mass defined in Eq.~(\ref{mudef0}) and $T$ is the temperature defined in Eq.~(\ref{Tdef}).

Formulating the linearized two-fluid equations in terms of the variation of an action as in Eq.~(\ref{S0c3}) allows us to develop \textit{variational} solutions of these equations by making an ansatz for the displacement fields $\bu_s(\br,t)$ and $\bu_n(\br,t)$.  This was done in Ref.~\cite{TaylorPRA05}, extending earlier work in Ref.~\cite{ZNG} for the two-fluid modes of a trapped Bose-condensed gas at finite temperatures.  
Our variational ansatz for each Cartesian component of the displacement fields is 
\bea u_{si} (\br,t) &=&  a_{si}f_{i}(\br)\cos \omega t,\nonumber\\
u_{ni} (\br,t)&=& a_{ni}g_{i}(\br)\cos \omega t . \label{RR} \eea
The constants $a_{si}$ and $a_{ni}$ are the variational
parameters.  With an ansatz of this form, the variational equations reduce to \bea \frac{\partial S^{(2)}}{\partial a_{si}} = 0, \; \frac{\partial S^{(2)}}{\partial a_{ni}} = 0.\eea Once some suitable ansatz is made for the functions $f_i(\br)$ and $g_i(\br)$ in Eq.~(\ref{RR}), these equations can be used to generate variational solutions of the two-fluid equations and the corresponding normal mode frequencies $\omega$. 

For gases confined in a harmonic trap, there exist
simple trial functions for $f_i(\br)$ and $g_{i}(\br)$ which are
sufficiently close to the exact solutions that good results for the mode frequencies $\omega$
are obtained by considering only a single expansion term as in
Eq.~(\ref{RR})~\cite{ZNG}.  The choice of ansatz for the displacement fields at finite
temperatures used in Ref.~\cite{TaylorPRA05} for the dipole and breathing modes are guided by the known exact hydrodynamic solutions at
$T=0$~\cite{Stringari96} and $T>T_c$~\cite{GriffinStringari,Bruun99}.  For the breathing mode, we use
\bea f_i(\br) = x_i, \; g_i(\br) = x_i.\label{bmansatz}\eea
For an isotropic trap, the \textit{breathing} mode in Eq.~(\ref{RR}) is described by $a_{si} \equiv a_s$ and $a_{ni} \equiv a_n$, in which case the displacement fields are given by \bea \bu_s(\br,t) = a_s \br \cos\omega t,\;\bu_n(\br,t) = a_n\br\cos\omega t.\label{bmansatzu}\eea  

The \textit{dipole} mode is characterized by displacements of the
centre-of-masses of the two fluids along one of the axes of the
harmonic trap, say the $z$ axis.  In this case, we use the following ansatz for the
displacement fields:
\bea f_z(\br) = a_s,\; g_z(\br) = a_n, \label{dipansatz}
\eea
where $a_s$ and $a_n$ describe the displacements of the centre-of-masses of the two fluids from the trap centre.  This ansatz describes a uniform displacement field,
\bea \bu_s(\br,t) = a_s\hat{\bz}\cos \omega t, \; \bu_n(\br,t) = a_n\hat{\bz}\cos \omega t.\label{dmansatzu}\eea

At $T=0$ where the normal fluid component vanishes, the ansatz used above for the breathing and dipole modes are exact solutions of the quantum hydrodynamic equations.  Similarly, for $T>T_c$, where the superfluid component vanishes, $\bu_n(\br,t) = a_n  \br\cos \omega t$ and $\bu_n(\br,t) = a_n\hat{\bz}\cos\omega t$ are both solutions of the collisional hydrodynamic equations~\cite{GriffinStringari,Bruun99}.  

We expect that the ansatz given above will be a good approximation to the exact solutions in the superfluid two-fluid region.  We note, however, that it is straightforward to improve the results presented in this paper by extending our variational ansatz using a generalized Rayleigh-Ritz expansion~\cite{ZNG}.  For the breathing mode, for instance, this would take the form \bea \bu_{s,n} = \sum_{j=0}^N a^{(j)}_{s,n}\br\; r^{2j}\cos\omega t.\label{bmansatzext}\eea 
In addition to improving our numerical results for the lowest breathing mode ($n=1,l=0$) frequency, this ansatz also allows us to solve for the higher-order ($n>1,l=0$) ``monopole" modes (see, for instance, Ref.~\cite{Pethickbook}).  

We note that the ansatz for our breathing mode in Eq.~(\ref{bmansatzu}) satisfies $\bnab(\bnab\cdot\bu)=0$.  Similarly, the dipole mode ansatz in Eq.~(\ref{dmansatzu}) satisfies $\bnab\cdot\bu = 0$.  We recall from our analysis in Section~\ref{isentropic} that the Landau two-fluid equations thus require the resulting in-phase \textit{breathing} mode to be locally isentropic (corresponding to $a_s = a_n$) only at unitarity, while the in-phase \textit{dipole} mode is locally isentropic everywhere.  In Section~\ref{bmu} we confirm that our variational solution of the in-phase \textit{breathing} mode is described by $a_s = a_n$ at unitarity (and \textit{only} at unitarity).  That $a_s = a_n$ for the in-phase \textit{dipole} mode is always correct has already been shown in Refs.~\cite{ZNG,TaylorPRA05}.

\section{Superfluid density and adiabatic compressibility of a uniform
Fermi gas at unitarity}
\label{NSRsec}

In later sections, we show that our variational solutions for the two-fluid dipole and breathing modes at unitarity require as input only two thermodynamic quantities: the superfluid density $\rho _s$ and the adiabatic compressibility $(\partial
\mu /\partial \rho )_s$.  In this section, we discuss the approximations used to evaluate these quantities.

The adiabatic compressibility $(\partial \mu /\partial \rho )_s$ can be
extracted from the equation of state for a uniform Fermi gas at unitarity.
For this purpose, we express the chemical potential and the entropy in terms
of dimensionless functions as a function of the reduced temperature, 
\begin{equation}
\mu =\frac{\epsilon _F\left( \rho \right)}{m} f_\mu \left[ T/T_F\left( \rho \right)
\right] ,  \label{mu}
\end{equation}
and 
\begin{equation}
s=\frac{\rho k_B}{m} f_s\left[ T/T_F\left( \rho \right) \right] ,  \label{entropy}
\end{equation}
where the dimensionless functions $f_\mu $ and $f_s$, and their derivatives
may be calculated numerically using the fluctuation theory discussed in Refs.~\cite{HLD,HDL2007}.

Using Eq.~(\ref{mu}), the compressibility is given by
\begin{equation}
\frac{\partial \mu }{\partial \rho }=\frac 23\frac{\epsilon _F\left( \rho
\right) }{m\rho} f_\mu +\frac{\epsilon _F\left( \rho \right)}{m} f_\mu ^{\prime }\frac{%
\partial \left[ T/T_F\left( \rho \right) \right] }{\partial \rho },\label{compressmicro}
\end{equation} where $f^{\prime} \equiv df/dT^{\prime}$ with $T^{\prime} \equiv T/T_F$.  
From the expression in Eq.~(\ref{entropy}), we see that keeping the entropy constant in evaluating Eq.~(\ref{compressmicro}) amounts to requiring that 
\begin{equation}
\frac{\partial \left[ T/T_F\left( \rho \right) \right] }{\partial \rho }=-%
\frac 1\rho \frac{f_s}{f_s^{\prime }}.
\end{equation}
We thus obtain 
\begin{equation}
\left( \frac{\partial \mu }{\partial \rho }\right) _s=\frac{\epsilon
_F\left( \rho \right) }{m \rho} \left[ \frac 23f_\mu -\frac{f_\mu ^{\prime }f_s}{%
f_s^{\prime }}\right].
\end{equation}
This quantity is straightforwardly evaluated using the values $f_\mu $ and $f_s$ obtained from the finite temperature equation of state of a uniform superfluid.

\begin{figure}
\begin{center}
\epsfig{file=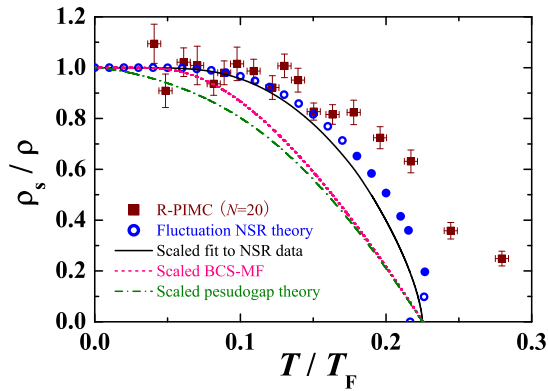, angle=0,width=0.40\textwidth}
\caption{(color online) Superfluid density fraction in a uniform Fermi gas at unitarity as a function of temperature.  The different theoretical predictions are discussed in the text. 
}
\label{nsfig}
\end{center}
\end{figure}
The determination of the superfluid density in the BCS-BEC crossover is more subtle. It has
been recently calculated for a uniform system including Gaussian NSR fluctuations~\cite{TaylorPRA06,TaylorPRA07}, and by Akkineni {\it et al.} using
path-integral Monte Carlo (PIMC) simulations~\cite{Trivedi06}. We
summarize these results in Fig.~\ref{nsfig}. Neither the NSR fluctuation or the PIMC calculations give results that are accurate near the
superfluid transition temperature $T_c$. The PIMC calculation suffers from
the negative-sign problem for fermions, and the results are thus restricted
to a small number of total particles $N=20$. The NSR-type Gaussian fluctuation theory, on the
other hand, suffers from a re-entrance problem close to $T_c$.  This problem first appears around $(k_Fa_s)^{-1} = -0.5$ on the BCS side and persists into the BEC side of unitarity~\cite{TaylorPRA07}.  This spurious first-order phase transition~\cite{TaylorPRA07} is due to the NSR Gaussian treatment of pairing fluctuations used to calculate $\Delta_0$ and $\mu$ self-consistently.  The problem is equivalent to one that arises in a self-consistent calculation of the condensate density and chemical potential close to $T_c$ in Bose gases using the Bogoliubov-Popov approximation (for further discussion  and references, see p.~34 of Shi and Griffin~\cite{Shi}).  However, as seen in Fig.~\ref{nsfig}, both the NSR and PIMC data are in good agreement at low temperatures.  

To overcome the lack of an accurate $\rho_s$ calculation near $T_c$, we use two different sets of data for the superfluid density in our calculation of the out-of-phase breathing and dipole modes: a fit to the NSR data from Ref.~\cite{TaylorPRA07} and a scaled BCS mean-field superfluid density.  A superfluid with a two-component order parameter (and a bosonic fluctuation spectrum) undergoes a second order phase transition with a superfluid density that varies as $\rho_s \propto (T_c -T)^{2/3}$ close to the transition temperature, independent of the interaction strength~\cite{statphys2}.  Our fit to the NSR fluctuation data thus assumes a curve of the form $(T_c - T)^{2/3}$ in the region $T>0.18T_F$.  This leads to the fitting curve $\rho_s/\rho = 4.51(0.237-T/T_F)^{2/3}$ for the high temperature data, where $T_c$ is
$0.237T_F$ in this fitting.  We scale the temperature dependence of this data using $T\rightarrow (0.225/0.2368)T$, so that the transition temperature for a uniform gas is $T_c \simeq 0.225T_F$, as given by NSR theory.  The final result is plotted in Fig.~\ref{nsfig} (``scaled fit to NSR data").  The original NSR data from Ref.~\cite{TaylorPRA07} is denoted by the blue circles (the data points used in the curve fitting are given by the filled circles).

The PIMC results for $\rho_s$ shown in Fig.~\ref{nsfig} have not been rescaled to the $T_c$ used for the other predictions.  Akkineni {\it et al.}~\cite{Trivedi06} have used finite-size scaling procedures to obtain a $T_c\simeq0.25T_F$.  One can ignore the PIMC data points above $0.25T_F$ and introduce a smooth extrapolation of the lower temperature points to vanish at $0.25T_F$.  When plotted in Fig.~\ref{nsfig} using a rescaled $T_c$ of $0.225T_F$, the resulting PIMC results are in fairly good agreement with our fitted NSR results.

The NSR-type theories developed in Refs.~\cite{HLD,HDL2007,TaylorPRA06,TaylorPRA07} includes the contributions from the BCS Fermi excitations plus the bosonic pairing fluctuations.  As discussed in Refs.~\cite{TaylorPRA06,TaylorPRA07}, one finds that the normal fluid density $\rho_n (=\rho-\rho_s)$ reduces precisely to the expected Landau formulas in both the BCS and BEC limits.  That is, the normal fluid is expressed in terms of Fermi excitations (BCS) or Bogoliubov-Popov Bose excitations (BEC), respectively.  Obtaining both limits correctly is very important in any acceptable theory of the superfluid density in the BCS-BEC crossover.  As noted above, however, even though our expression for the superfluid density reduces to the Landau expression on the BEC side of unitarity, our results are still unreliable near $T_c$ because the spectrum of Bogoliubov-Popov excitations that determines $\rho_s$ is evaluated using the values of $\Delta_0(T)$ and $\mu(T)$ determined self-consistently in the Gaussian NSR theory.   The self-consistent determination of $\Delta_0(T)$ and $\mu(T)$ in this approximation is equivalent to the calculation of the condensate density $n_c(T)$ and chemical potential $\mu(T)$ for a Bose gas with a Bogoliubov-Popov excitation spectrum~\cite{TaylorPRA06}.  It is well-known (see Shi and Griffin~\cite{Shi}) that the latter problem predicts a spurious first-order phase transition because one is trying to determine the condensate depletion self-consistently from the thermal excitation of collective modes with a spectrum that depends on the condensate fraction. 

As we have noted, the NSR-type treatment of fluctuations appears to give excellent results for the thermodynamic functions in the BCS-BEC crossover when one compares them with \textit{ab-initio} calculations.  The NSR theory does have a problem near $T_c$ near unitarity and on the BEC side of the crossover as a result of only considering Gaussian fluctuations.  However, we consider it the best available theory for the superfluid density $\rho_s$ in the BCS-BEC crossover at the present time.

In addition to the fitted NSR data for $\rho_s$, we also use a scaled mean-field BCS superfluid density fraction.  This is obtained by
a linear compression of the horizontal axis of the BCS superfluid density, 
\begin{equation}
\rho _s^{scaled}[T]=\rho _s^{BCS}[\frac{T_c^{NSR}}{T_c^{BCS}}T].\label{nsscaled}
\end{equation}
Here $T_c^{NSR}\simeq 0.225T_F$ and $T_c^{BCS}\simeq 0.497T_F$ are the
transition temperatures of the uniform Fermi gas given by the NSR
fluctuation theory~\cite{HLD,TaylorPRA07} and the mean-field BCS theory, respectively.
This data is shown in Fig.~\ref{nsfig} by a dotted line.  While calculations~\cite{TaylorPRA07} show that the BCS result for $\rho_s$ is only a good description of the superfluid density on the BCS side of resonance when $(k_Fa_s)^{-1}\lesssim -0.5$, much of the effect of ``beyond mean-field fluctuations" is included by the scaling of $T_c$ in Eq.~(\ref{nsscaled}).  It is well-known that near $T_c$ a mean-field BCS type theory has a $\rho_s\propto (T_c -T)$ outside the region where fluctuations are important.

He, Chien, Chen, and Levin~\cite{Levin07b} have also recently calculated the superfluid density in the BCS-BEC crossover using the pseudogap theory~\cite{Levinreview}.  For comparison, in Fig.~\ref{nsfig}, we also plot the pseudogap result for $\rho_s$ at unitarity.  This is obtained by evaluating the expression given in Eq.~70 of Ref.~\cite{Levin07b} using values of the gap and chemical potential
obtained by solving Eqs.~(34)-(36) of Ref.~\cite{Levin07b}.  The pseudogap expression for $\rho_s$ is given by $\rho_s = \Delta^2_{sc}/\Delta^2 \rho^{BCS}_s(\Delta)$, where $\rho^{BCS}_s(\Delta)$ is the BCS mean-field superfluid density with a modified pairing gap $\Delta$.  The effective gap is now renormalized to $\Delta = (\Delta^2_{sc} + \Delta^2_{pg})^{1/2}$, where $\Delta_{pg}$ is a temperature-dependent pseudogap describing the effect of bound pairs of the Fermi excitations.  As shown in Fig.~\ref{nsfig}, the pseudogap $\rho_s$ is very similar to the rescaled BCS result at higher temperatures.  At low temperatures, the prefactor $\Delta^2_{sc}/\Delta^2$ in the pseudogap expression for $\rho_s$ leads to a normal fluid density $\rho_n = \rho-\rho_s \propto T^{3/2}$ (like an ideal Bose gas of molecules).  We refer to Ref.~\cite{Levinreview} for more details.

It is still not clear how to assess the treatment of bosonic pairing fluctuations used in the pseudogap calculation~\cite{Levin07b,Levinreview}.  One indication of what it misses is to consider the BEC limit of the crossover, in which case the superfluid density predicted by Ref.~\cite{Levin07b} reduces to the condensate fraction of a noninteracting Bose gas of molecules, rather than the superfluid density for a gas of Bogoliubov excitations.  This suggests that the pseudogap theory does not have any self-consistency problem near $T_c$ because it leaves out the interactions between bound pairs (equivalent to working with an ideal Bose gas).  In future work, we will give a more detailed comparison between the NSR-type fluctuation theory we use and the renormalized mean-field BCS theory involving a pseudogap.

\section{Breathing mode frequencies}
\label{bmf}

The variational equations for the breathing mode frequencies using the ansatz in Eq.~(\ref{bmansatz}) are~\cite{TaylorPRA05}
\bea \tilde{M}_{si}\omega^2 a_{si} =\frac{1}{2}\sum_j\Big[(k^s_{ij} + k^s_{ji})a_{sj}+ 2k^{sn}_{ij}a_{nj}\Big],
 \label{monopoles}\eea
\bea \tilde{M}_{ni}\omega^2 a_{ni}=\frac{1}{2}\sum_j\Big[(k^n_{ij}+ k^n_{ji})a_{nj} + 2k^{sn}_{ji}a_{sj}\Big].
 \label{monopolen}\eea
Here the ``mass moments" $\tilde{M}_i$ are defined by
\bea \tilde{M}_{si} \equiv \int d\br\;\rho_{s0}x_i^2 ,\;\;\;\;\tilde{M}_{ni} \equiv \int
d\br\;\rho_{n0}x_i^2, \label{breathingmasses}\eea
and the ``spring constants" $k^s_{ij}, k^{n}_{ij}$, and $k^{sn}_{ij}$ are 
\bea k^{s}_{ij} &=& \int d\br\;\left[\left(\frac{\partial
\mu}{\partial
\rho}\right)_{\!s}\frac{\partial(\rho_{s0}x_i)}{\partial
x_i}\frac{\partial(\rho_{s0}x_j)}{\partial
x_j}\right], \label{ksm} \eea
\bea k^n_{ij} &=& \int d\br\;\Bigg[\left(\frac{\partial
\mu}{\partial
\rho}\right)_{\!s}\frac{\partial(\rho_{n0}x_i)}{\partial
x_i}\frac{\partial(\rho _{n0}x_j)}{\partial x_j} \nonumber\\&&+
2\left(\frac{\partial T}{\partial
\rho}\right)_{\!s}\frac{\partial(\rho_{n0}x_i)}{\partial x_i}
\frac{\partial(s_{0}x_j)}{\partial x_j}\nonumber\\
&&+
\left(\frac{\partial T}{\partial
s}\right)_{\!\rho}\frac{\partial(s_{0}x_i)}{\partial
x_i}\frac{\partial(s_{0}
x_j)}{\partial x_j}\Bigg], \label{knm}\eea
and \bea 
k^{sn}_{ij} &=&\int d\br\;\Bigg[\left(\frac{\partial
\mu}{\partial
\rho}\right)_{\!s}\frac{\partial(\rho_{s0}x_i)}{\partial
x_i}\frac{\partial(\rho_{n0}x_j)}{\partial x_j} \nonumber\\&&+
\left(\frac{\partial T}{\partial
\rho}\right)_{\!s}\frac{\partial(\rho_{s0}x_i)}{\partial
x_i}\frac{\partial(s_0x_j)}{\partial x_j}\Bigg].\label{ksnm}\eea 

To solve for the breathing modes (see Section~\ref{bmu}), it is useful to rewrite the above equations.  We define the following new coefficients involving the spring constants:
\bea K^{s}_{ij}\equiv 2k^s_{ij} + 2k^{sn}_{ij}\label{Omegas}\eea
and
\bea K^n_{ij} \equiv  k^{n}_{ij} + k^{n}_{ji} + 2k^{sn}_{ji}.\label{Omegan}\eea
Adding the two equations for the breathing modes in Eqs.~(\ref{monopoles}) and (\ref{monopolen}), we obtain
\bea \omega^2(\tilde{M}_{si}a_{si} + \tilde{M}_{ni}a_{ni})
&=&\frac{1}{2}\sum_j\left(K^n_{ji}a_{nj} + K^s_{ji}a_{sj}\right).\nonumber\\ \label{bmeqn1}\eea
Furthermore, dividing Eqs.~(\ref{monopoles}) and (\ref{monopolen}) by $\tilde{M}_{si}$ and $\tilde{M}_{ni}$ respectively, and subtracting one from the other, we obtain
\bea \omega^2(a_{si} - a_{ni})&=&\frac{1}{2}\sum_j\Bigg[\left(\frac{2k^s_{ij}}{\tilde{M}_{si}} + \frac{2k^s_{ij}}{\tilde{M}_{ni}} - \frac{K^s_{ji}}{\tilde{M}_{ni}}\right)\nonumber\\ &&\!\!\!\!\!\!\!\!\!\!\!\!\!\!\!\!\!\!\!\times (a_{sj} - a_{nj}) + \left(\frac{K^s_{ij}}{\tilde{M}_{si}} -  \frac{K^n_{ij}}{\tilde{M}_{ni}}\right)a_{nj}\Bigg].  \label{bmeqn2}\eea

After some rearranging, we can write the coefficients defined in Eqs.~(\ref{Omegas}) and (\ref{Omegan}) as
\bea K^s_{ij}&=&2\int d\br\frac{\partial \rho_{s0}x_i}{\partial x_i}\Bigg[x_j\frac{{\cal{D}}\mu}{{\cal{D}}x_j}\! +\!\rho_0\left(\frac{\partial \mu}{\partial\rho}\right)_{\!s}\! +\! s_0\left(\frac{\partial \mu}{\partial s}\right)_{\!\rho}\Bigg]\nonumber\\  \label{combo1}\eea and
\bea K^n_{ij}&=&2\int d\br\frac{\partial \rho_{n0}x_i}{\partial x_i}\Bigg[x_j\frac{{\cal{D}}\mu}{{\cal{D}}x_j}\!+ \!\rho_0\left(\frac{\partial \mu}{\partial\rho}\right)_{\!s}\!+\!s_0\left(\frac{\partial \mu}{\partial s}\right)_{\!\rho}\Bigg]\nonumber\\&+&2\int d\br\frac{\partial s_{0}x_i}{\partial x_i}\Bigg[x_j\frac{{\cal{D}}T}{{\cal{D}}x_j}\!+\!\rho_0\left(\frac{\partial T}{\partial\rho}\right)_{\!s}\! +\! s_0\left(\frac{\partial T}{\partial s}\right)_{\!\rho}\Bigg].\nonumber\\  \label{combo4}\eea
Here we have defined [not to be confused with the Lagrangian derivative defined in Eq.~(\ref{Lder})]
\bea \frac{{\cal{{\cal{D}}}}\mu}{{\cal{{\cal{D}}}}x_j}\equiv \left(\frac{\partial\mu}{\partial\rho}\right)_{\!s}\frac{\partial\rho_0}{\partial x_j} + \left(\frac{\partial\mu}{\partial s}\right)_{\!\rho}\frac{\partial s_0}{\partial x_j},\label{Dmu}\eea and
\bea \frac{{\cal{D}} T}{{\cal{D}}x_j}\equiv \left(\frac{\partial T}{\partial\rho}\right)_{\!s}\frac{\partial\rho_0}{\partial x_j} + \left(\frac{\partial T}{\partial s}\right)_{\!\rho}\frac{\partial s_0}{\partial x_j}.\label{DT}\eea
In writing down these equations, we also have made use of the Maxwell relation given by Eq.~(\ref{maxwell}).   

We next proceed to show that the expressions given in Eqs.~(\ref{combo1}) and (\ref{combo4}) can be written in terms $\rho_{s0}$, $\rho_{n0}$, and the two thermodynamic derivatives, $(\partial P/\partial\rho)_{s}$, and $(\partial P/\partial s)_{\rho}$. 
To handle the derivatives ${\cal{D}}(\mu, T)/{\cal{D}}x_j$, we note that the gradient of the equilibrium chemical potential $\mu_0$ can be written as [using Eq.~(\ref{mudef0})]
\bea \bnab\mu_0 =\left(\frac{\partial \mu}{\partial \rho}\right)_{\!s}\bnab\rho_0
+\left(\frac{\partial \mu}{\partial s}\right)_{\rho}\bnab s_0 +\bnab
V_{\mathrm{ext}}. \label{bnabmu}\eea  
Recall that in equilibrium, both the temperature and the chemical potential are spatially uniform ($\bnab\mu_0 =\bnab T_0 = 0$).  Thus, for a harmonic trapping potential given by Eq.~(\ref{trap}), Eq.~(\ref{bnabmu}) reduces to 
\bea\left(\frac{\partial
\mu}{\partial \rho}\right)_{\!s} 
\frac{\partial \rho_0}{\partial x_j} + \left(\frac{\partial
\mu}{\partial s}\right)_{\!\rho}\frac{\partial s_0}{\partial x_j}&=&-\frac{\partial}{\partial 
x_j}V_{\mathrm{ext}}\nonumber\\&=&
-\omega_j^2x_j, \label{useful3} \eea  where $\omega_j$ is the trap frequency along the $x_j$-axis. 
Using the results in Eqs.~(\ref{dTdx}) and (\ref{useful3}), Eqs.~(\ref{Dmu}) and (\ref{DT}) simplify to
\bea \frac{{\cal{D}}\mu}{{\cal{D}}x_j}= -\omega^2_jx_j\label{identity1}\eea
and
\bea  \frac{{\cal{D}} T}{{\cal{D}} x_j}=0.\label{ident}\eea  

Using Eqs.~(\ref{identity2}), (\ref{identity3}), (\ref{identity1}), and (\ref{ident}), and integrating by parts, the new spring constants $K^s_{ij}$ and $K^n_{ij}$ in Eqs.~(\ref{combo1}) and (\ref{combo4}) reduce to                      
\bea K^s_{ij} &=&2\int d\br\;\rho_{s0}x_i\left[2\delta_{ij}\omega^2_i x_i - \frac{\partial}{\partial x_i}\left(\frac{\partial P}{\partial\rho}\right)_{\!s}\right], \label{springidentity3}\eea  and
\bea K^n_{ij} &=&2\int d\br\Big\{\rho_{n0}x_i\left[2\delta_{ij}\omega^2_i x_i- \frac{\partial}{\partial x_i}\left(\frac{\partial P}{\partial\rho}\right)_{\!s}\right] \nonumber\\&&-s_0x_i\frac{\partial}{\partial x_i}\left(\frac{\partial P}{\partial s}\right)_{\!\rho}\Big\}. 
\label{springidentity4}\eea 

In summary, we have reduced the algebraic equations for the breathing modes given by Eqs.~(\ref{monopoles}) and (\ref{monopolen}) to the set of equations given by Eqs.~(\ref{bmeqn1}) and (\ref{bmeqn2}), with the simpler spring constants given by Eqs.~(\ref{springidentity3}) and (\ref{springidentity4}).  Unlike the original spring constants defined in Eqs.~(\ref{ksm})-(\ref{ksnm}), these new spring constants $K^{s,n}_{ij}$ only involve derivatives of the pressure.   

Making use of the special properties of universal thermodynamics at unitarity, the spring constants $K^s_{ij}$ and $K^n_{ij}$ given in Eqs.~(\ref{springidentity3}) and (\ref{springidentity4}) reduce to simple expressions that involve only the mass moments $\tilde{M}_{si}$, $\tilde{M}_{ni}$ and the trap frequencies $\omega_i$.  
Using Eqs.~(\ref{identity4}) and (\ref{identity5}) in Eqs.~(\ref{springidentity3}) and (\ref{springidentity4}), we obtain:
\bea K^s_{ij} = 2\tilde{M}_{si}(2\delta_{ij} + 2/3)\omega^2_i\label{rel1}\eea
and
\bea K^n_{ij} = 2\tilde{M}_{ni}(2\delta_{ij} + 2/3)\omega^2_i.\label{rel2}\eea
These results (valid at unitarity) will be used in the next section.

\section{Breathing modes at unitarity}
\label{bmu}

Using Eqs.~(\ref{rel1}) and (\ref{rel2}) in Eq.~(\ref{bmeqn1}) and (\ref{bmeqn2}), the variational equations reduce to
\bea \lefteqn{\omega^2\left(\tilde{M}_{si}a_{si} + \tilde{M}_{ni}a_{ni}\right)
=}&&\nonumber\\ &&\sum_j(2\delta_{ij} + 2/3)\omega^2_j\left(\tilde{M}_{sj} a_{sj} +\tilde{M}_{nj}a_{nj}\right),\label{bmeqn1u}\eea
and
\bea \lefteqn{\omega^2(a_{si} - a_{ni})=}&&\nonumber\\&&
\sum_j\left[\frac{k^s_{ij}}{\tilde{M}_{si}} + \frac{k^s_{ij}}{\tilde{M}_{ni}} - 
\frac{\tilde{M}_{sj}}{\tilde{M}_{ni}}\left(2\delta_{ij} + 2/3\right)\omega^2_j\right](a_{sj} - a_{nj}). \nonumber\\ \label{bmeqn2u}\eea  
These equations for the variational parameters $a_s,a_n$ will be used to determine the hydrodynamic breathing modes at unitarity.

By inspection, one immediately sees that Eq.~(\ref{bmeqn2u}) has a solution given by
\bea a_{si} = a_{ni}. \label{bmsol1}\eea This solution corresponds to a solution of the Landau two-fluid equations of the form
\bea \bv_s(\br,t) = \bv_n(\br,t). \eea  This describes the expected locally isentropic (also isothermal) breathing mode at unitarity.  In Section~\ref{bmuip}, we show that the frequency of this mode is independent of temperature, as argued by Thomas~\textit{et al.}~\cite{Thomas05}.  Substituting the in-phase solution Eq.~(\ref{bmsol1}) into Eq.~(\ref{bmeqn1u}), the latter reduces to
\bea\tilde{M}_i\omega^2a_i
&=&\sum_j\tilde{M}_j\omega^2_j\left(2\delta_{ij} + 2/3\right)a_j,\label{bmeqn1u2}\eea
where $\tilde{M}_i \equiv \tilde{M}_{si} + \tilde{M}_{ni}$ and $a_{si} = a_{ni} \equiv a_i$.  
From the definition of the spring constants $K^s_{ij}$ and $K^n_{ij}$ in Eqs.~(\ref{Omegas}) and (\ref{Omegan}), one sees that $K^s_{ij} + K^n_{ij} = K^s_{ji} + K^n_{ji}$.  Applying this result, we see that the expressions in Eqs.~(\ref{rel1}) and (\ref{rel2}) imply (valid at unitarity)
\bea \tilde{M}_i\omega^2_i = \tilde{M}_j\omega^2_j\eea for all coordinates, $i,j = x,y,z$.  
Making use of this result in the right-hand side of Eq.~(\ref{bmeqn1u2}), it reduces to 
\bea
\omega^2_{1B} a_{i}=2\omega_i^2a_{i} + \frac{2}{3}\omega_i^2\sum_ja_{j},\label{mode1}\eea
where $\omega_{1B}$ denotes the frequency of the in-phase breathing mode. 
Equations~(\ref{bmsol1}) and (\ref{mode1}) describe the in-phase oscillation of the normal and superfluid components corresponding to $\bv_s = \bv_n$.
We discuss the solutions of Eq.~(\ref{mode1}) below.  

In addition to the in-phase solution in Eq.~(\ref{bmsol1}), there is an out-of-phase solution corresponding to
\bea \tilde{M}_{si}a_{si} + \tilde{M}_{ni}a_{ni} = 0,\label{bmsol2}\eea which satisfies Eq.~(\ref{bmeqn1u}).  
Substituting this out-of-phase solution into Eq.~(\ref{bmeqn2u}), we find a closed equation for the $a_{si}$ parameters and the frequency $\omega_{2B}$ of the out-of-phase breathing mode at unitarity, namely
\bea \omega^2_{2B}a_{si}=\!
\sum_j\!\!\frac{\tilde{M}_{sj}}{\tilde{M}_{si}}\!\!\left[\!\frac{k^s_{ij}}{\tilde{M}_{rj}} \!-\!\frac{\tilde{M}_{ri}\tilde{M}_{sj}}{\tilde{M}_{rj}\tilde{M}_{ni}}\left(2\delta_{ij} \!+\! 2/3\right)\omega^2_j\right]a_{sj}. \nonumber\\ \label{mode2}\eea
Here we have defined the reduced mass moment $\tilde{M}_{ri}$ as
\bea \tilde{M}_{ri} \equiv  \frac{\tilde{M}_{si}\tilde{M}_{ni}}{\tilde{M}_{si} + \tilde{M}_{ni}}.\eea

\subsection{In-phase mode at unitarity}\index{\footnote{}}
\label{bmuip}
The in-phase mode given by Eq.~(\ref{mode1}) is a normal mode of the Landau two-fluid equations at unitarity, valid at all temperatures.  It is equivalent to Eq.~(3) in
Ref.~\cite{Stringari05} for the zero-temperature breathing mode frequency of a trapped Fermi gas at unitarity, assuming a polytropic equation of state~\cite{Trentoreview} $\mu(\rho) = \rho^{\gamma}$ with polytropic exponent $\gamma = 2/3$.
For an axisymmetric trap, $\omega_x = \omega_y \equiv \omega_{\bot}$, the axial and longitudinal breathing modes are characterized by solutions of the
form $a_x = a_y$.  In this case, the solution of Eq.~(\ref{mode1}) is well known~\cite{Stringari05},
\bea
\omega^2_{1B}  = \frac{5}{3}\omega_{\bot}^2\! +\!
\frac{4}{3}\omega_z^2\pm \frac{1}{6}\sqrt{\left(10\omega_{\bot}^2 \!-\! 8\omega_z^2\right)^2\! +\! 32\omega_z^2\omega_{\bot}^2}.
\label{Stringarieosu} \eea 
We conclude that the frequency of the in-phase two-fluid hydrodynamic breathing mode at unitarity is independent of temperature and equal to the zero temperature value.  We further note that for an isotropic trap $(\omega_x =\omega_y=\omega_z \equiv \omega_0)$, we have $a_x = a_y = a_z$, and the in-phase mode frequency in Eq.~(\ref{Stringarieosu}) reduces to 
\bea \omega_{1B} = 2\omega_0.\label{mode1iso}\eea
Castin~\cite{Castin} has argued that there is an \textit{exact} eigenstate of the isotropic trap Hamiltonian such that all atoms move with velocity $\bv(r,t) = a\br\cos \omega t$ at all temperatures, giving rise to a temperature independent mode with frequency $\omega = 2\omega_0$.  It is reassuring that two-fluid hydrodynamics gives a result in agreement with this prediction~\cite{note2}.  The temperature independence of the in-phase breathing mode is also consistent with the results of the direct numerical solution of the Landau two-fluid equations reported in Ref.~\cite{Levin07}.

The fact that the in-phase breathing mode frequencies are independent of temperature is a consequence of the special thermodynamic properties at unitarity, and is not expected to hold away from unitarity.  In typical experiments where the trap is highly anisotropic ($\omega_z\ll \omega_{\perp}$), the radial hydrodynamic breathing mode frequency [given by the upper branch of Eq.~(\ref{Stringarieosu})] is well-approximated by
\bea \omega_{1B} \simeq \sqrt{10/3}\omega_{\bot}.\label{hd}\eea
The predicted temperature independence of the in-phase breathing mode frequency is consistent with the experimental results of Thomas and coworkers~\cite{Thomas05}. They found only a small difference (a few percent) between the measured radial breathing mode frequency and Eq.~(\ref{hd}) over a large temperature range, including well into the normal phase.

\subsection{Out-of-phase mode at unitarity}
\label{outofphasebmsec}

We now discuss the out-of-phase breathing mode at unitarity.  In the special limit of an isotropic trap, $a_{si} \rightarrow a_s$ and $\tilde{M}_{si}\rightarrow M^B_s/3$, $\tilde{M}_{ni}\rightarrow M^B_n/3$, where
\bea M^B_{s} \equiv \int d\br\;\rho_{s0}(r) r^2\label{Mstildeiso}\eea
and
\bea M^B_{n} \equiv \int d\br\;\rho_{n0}(r) r^2\label{Mntildeiso}.\eea
Using these identities, Eq.~(\ref{mode2}) simplifies to
\bea \omega^2_{2B} = \frac{k^B_{s}}{M^B_r} - 4\frac{M^B_{s}}{M^B_n}\omega^2_0,\label{mode2iso}\eea 
where we have defined the breathing mode spring constant,
\bea k^B_s &\equiv& 3\sum_j k_{s,ij}=\sum_{i,j}k_{s,ij}\nonumber\\
&=&\int d\br\;\left(\frac{\partial\mu}{\partial \rho}\right)_{\!s}\left[\bnab\cdot\left(\br \rho_{s0}(\br)\right)\right]^2.\label{ksiso}\eea
The reduced mass moment is now given by $M^B_r \equiv M^B_sM^B_n/(M^B_s + M^B_n)$. 
As follows from Eq.~(\ref{bmsol2}), this out-of-phase mode corresponds to the following eigenvector:
\bea M^B_{s}a_{s} + M^B_{n}a_{n} = 0.\label{bmsol2iso}\eea 

We now present numerical results for the out-of-phase breathing mode. From Eq.~(\ref{mode2iso}), we see that only two thermodynamic functions enter in the evaluation of the mode frequency: the superfluid density $\rho _s(r)$ and the adiabatic compressibility $(\partial \mu /\partial \rho )_s(r)$.  The calculation of these quantities in a uniform gas within an NSR-type formalism is discussed in Section~\ref{NSRsec}.  We use a local density approximation (LDA) to calculate the local superfluid density and compressibility in a trapped Fermi gas.  

The local density approximation in an isotropic harmonic trap ($\omega_0$) amounts to determining the global chemical
potential $\mu$ from the local equilibrium condition 
\begin{equation}
\mu =\mu _{\hom }\left[ \rho (r),T/T_F(\rho )\right] +\omega _0^2r^2/2.\label{LDA}
\end{equation}
Here the local reduced temperature $T/T_F(\rho)$ depends on the local mass density $\rho
(r)$.  Eq.~(\ref{LDA}) is solved for the the density profile $\rho(r)$, subject to the constraint
\begin{equation}
\int d{\bf r}\rho (r)=Nm.\label{Nconserve}
\end{equation}
To solve for $\rho(r)$ using LDA, for a given temperature 
$T$, we tabulate the local chemical potential as a function of the mass
density using Eq. (\ref{mu}).  With an initial guess of the global
chemical potential, we determine the local chemical potential from the local
equilibrium condition in Eq.~(\ref{LDA}), and invert it in tabular form to find the mass
density. The global chemical potential is then adjusted slightly to enforce
the number conservation requirement in Eq.~(\ref{Nconserve}), giving a better estimate for the next iterative
step. 

In a harmonic trap, it is convenient to use the trap units, where $%
m=k_B=\hbar =\omega _0=1$, {\it i.e.}, we take the characteristic harmonic
oscillator length $a_{ho}=\sqrt{\hbar /m\omega _0}$ and the characteristic
level spacing $\hbar \omega _0$ as the units of the length and energy,
respectively.  We use the Fermi energy $%
E_F=(3N)^{1/3}\hbar \omega _0$ and the corresponding temperature $T_F=E_F/k_B
$ of an ideal Fermi gas to characterize the energy scale and the temperature
scale, where $N$ is the total number of atoms.  The distance and
the mass density are conveniently given in units of the Thomas-Fermi radius $R_{TF}=(24N)^{1/6}a_{ho}$ for an ideal Fermi gas, and the mass density at the centre of the trap, $%
\rho _{TF}=(24N)^{1/2}/(3\pi ^2)ma_{ho}^{-3}$, respectively.

\begin{figure}
\begin{center}
\epsfig{file=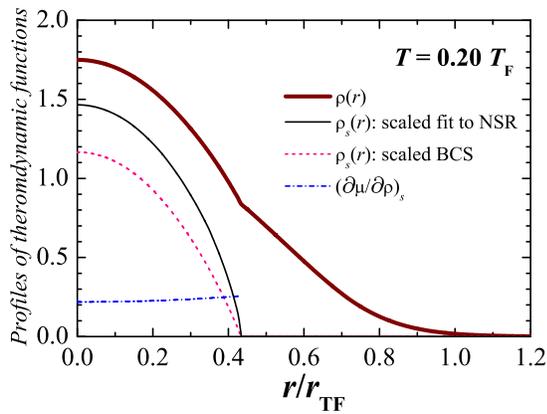, angle=0,width=0.40\textwidth}
\caption{(color online) Profiles of several thermodynamic functions at
temperature $T=0.20T_F$, where $T_F$ is the Fermi temperature of a trapped
ideal Fermi gas. Here we have used the results given by the NSR-type Gaussian fluctuation theory~\cite{HLD}.  The superfluid transition temperature is $T_c\simeq 0.27T_F$. The adiabatic compressibility is only needed in the superfluid
region of the trap.}
\label{thermofig}
\end{center}
\end{figure}

In Fig.~\ref{thermofig} we plot the profiles for the total mass density,
the superfluid mass densities using the results in Fig.~\ref{nsfig}, and the adiabatic compressibility. The total mass
density shows a bi-modal distribution, as expected from the general
universal argument \cite{Ho04}. The superfluid densities
drop to zero steeply at the superfluid-normal interface in the trap.

Having calculated the local adiabatic compressibility and the superfluid and total mass density profiles, it is straightforward to evaluate
the mass moments and the spring constant that enter the out-of-phase breathing mode frequency.

\begin{figure}
\begin{center}
\epsfig{file=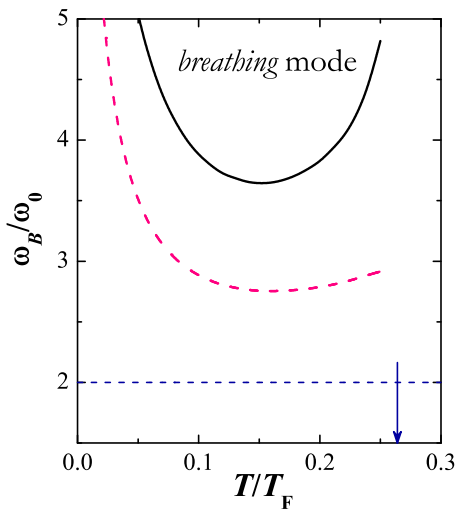, angle=0,width=6cm}
\caption{The out-of-phase breathing mode at unitarity as a function of temperature for an isotropic trap using an improved NSR calculation for the equation of state~\cite{HLD}.  The frequencies obtained using the fitted NSR $\rho_s$ data and the scaled BCS mean-field data are given by the solid and dashed lines, respectively (see Fig.~\ref{nsfig}). The arrow indicates the superfluid transition temperature, $T_c\simeq 0.27T_F$.}
\label{bmfig}
\end{center}
\end{figure}

The frequency $\omega _{2B}$ is plotted in Fig.~\ref{bmfig} using two approximations for the superfluid density, as a function of temperature. 
One immediately sees that the out-of-phase breathing mode frequency is quite sensitive to the superfluid density in a uniform gas (using the LDA).  This underlines the importance of calculating $\rho_s$ with better accuracy as input into our variational theory.  However, the qualitative features of the temperature dependence of $\omega_{2B}$ are similar for both the fitted NSR and scaled mean-field BCS data for $\rho_s$.  Namely, the frequency increases rapidly at low temperatures and decreases with increasing temperature, before increasing again as $T_c$ is approached.  In both cases, the frequency of the out-of-phase breathing mode is larger than the in-phase breathing mode  
$\omega_{1B} = 2\omega_0$.  These features are quite different from the results of He \textit{et al.}~\cite{Levin07}.  They find the out-of-phase breathing mode frequency starts \textit{below} the in-phase mode frequency at low temperatures, and increases monotonically as $T$ approaches $T_c$.

In Appendix~\ref{appendixD}, we argue that the divergence of $\omega_{2B}$ as $T\rightarrow 0$ is not an artifact of the local density approximation (LDA) we use to evaluate the coefficients in Eq.~(\ref{mode2iso}).  We compare the results of a mean-field LDA calculation directly with the results obtained by self-consistently solving the Bogoliubov-de Gennes equations and find excellent agreement.  The discussion in Appendix~\ref{appendixD} shows that LDA is not the source of any significant error at low temperatures.  

The increase of $\omega_{2B}$ as $T\rightarrow 0$ can be understood within our variational formalism as follows.  As emphasized in Ref.~\cite{TaylorPRA05}, our variational solutions of the two-fluid equations describe two coupled harmonic oscillators with effective masses given by the mass moments for the mode in question [for the breathing mode, these are given by Eqs.~(\ref{Mstildeiso}) and (\ref{Mntildeiso})].  As $T\rightarrow 0$, the mass of the normal fluid ``oscillator" goes to zero.  As with two coupled harmonic oscillators, in this limit, the small (normal fluid) mass executes a high frequency (and large amplitude) oscillation about the heavy (superfluid) mass, which is essentially static.  We should also caution that at low but finite $T$, the Landau two-fluid equations are no longer valid because local equilibrium cannot be established.  

The high-temperature ($T\rightarrow T_c$) behavior of $\omega_{2B}$ is discussed in Appendix~\ref{appendixC} based on the analytic expression in the BCS approximation for the superfluid density in a trapped gas~\cite{Baranov}.

\section{Dipole modes}
\label{secdipole}
The dipole modes discussed in Ref.~\cite{TaylorPRA05} are characterized by the uniform displacement fields given by Eq.~(\ref{dipansatz}).  
Inserting this ansatz into the action given in Eq.~(\ref{S0c3}) and taking its variation, one finds an in-phase oscillation (generalized Kohn mode) with $a_{s} = a_{n}$ and frequency $\omega_{1D} = \omega_z$ given by the trap frequency $\omega_{z}$ along the $z$-axis.  In addition, there is an out-of-phase mode corresponding to the solution $M_sa_s + M_na_n=0$.  The frequency of this mode is given by~\cite{TaylorPRA05}
\bea \omega^2_{2D} = \omega^2_z - \frac{k^D_{sn}}{M^D_r}.\label{secondsounddip}\eea
Here $M^D_r = M^D_sM^D_n/(M^D_s + M^D_n)$ is the reduced mass of the superfluid and normal fluid components, with \bea M^D_s \equiv \int d\br\;\rho_{s0}(\br) \eea and \bea M^D_n \equiv \int d\br\;\rho_{n0}(\br) \eea giving the masses of the superfluid and normal fluids.  The spring constant $k^D_{sn}$ in Eq.~(\ref{secondsounddip}) is defined as~\cite{TaylorPRA05}
\bea k^D_{sn} \equiv \int d\br\;\left[\left(\frac{\partial\mu}{\partial \rho}\right)_{\!s}\frac{\partial\rho_{n0}}{\partial z} + \left(\frac{\partial T}{\partial \rho}\right)_{\!s}\frac{\partial s_0}{\partial z}\right]\frac{\partial\rho_{s0}}{\partial z}.\label{ksndip}\eea
This is the analogue of the corresponding spring constant $k^{sn}_{ij}$ for the breathing mode, defined in Eq.~(\ref{ksnm}).   
\begin{figure}
\begin{center}
\epsfig{file=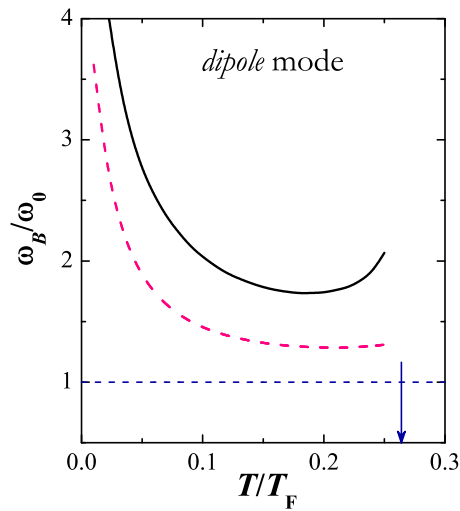, angle=0,width=6cm}
\caption{The frequency of the out-of-phase dipole mode at unitarity in an isotropic trap as a function of temperature.  See caption of Fig.~\ref{bmfig}}
\label{dmfig}
\end{center}
\end{figure}
We note that $k^D_{sn}$ is the negative of the analogous spring constant $k_{sn}$ in Eq.~(79) of Ref.~\cite{TaylorPRA05}.

It is convenient to write the out-of-phase mode frequency in Eq.~(\ref{secondsounddip}) in terms of a simpler spring constant $k^D_{s}$, which only involves the isentropic compressibility $(\partial\mu/\partial\rho)_{\!s}$.  Using Eqs.~(\ref{maxwell}) and (\ref{useful3}) in Eq.~(\ref{ksndip}), we find 
\bea k^D_{sn} &=& -\int d\br\;\left[\omega^2_z z + \left(\frac{\partial \mu}{\partial \rho}\right)_{\!s}\frac{\partial\rho_{s0}}{\partial z}\right]\frac{\partial\rho_{s0}}{\partial z}\nonumber\\&\equiv&M^D_s\omega^2_z - k^D_s,\eea
where 
\bea k^D_s \equiv  \int d\br\;\left(\frac{\partial \mu}{\partial \rho}\right)_{\!s}\left(\frac{\partial\rho_{s0}}{\partial z}\right)^2.\eea
Using these results in Eq.~(\ref{secondsounddip}), we find it reduces to
\bea
\omega^2_{2D} =
 \frac{k_{s}}{M_r}-\frac{M_s}{M_n}\omega^2_z. \label{secondsounddip2} \eea

This frequency of the out-of-phase dipole mode in an isotropic trap ($\omega_z = \omega_0$) is plotted in Fig.~\ref{dmfig} as a function of temperature.  As in Fig.~\ref{bmfig}, we compare the results obtained using the superfluid density given by NSR theory and the BCS mean-field approximation (see Fig.~\ref{nsfig}).  
We note that the expression given in Eq.~(\ref{secondsounddip2}) for the frequency of the out-of-phase dipole mode is very similar to the formula for the frequency of the out-of-phase isotropic breathing mode at unitarity given in Eq.~(\ref{mode2iso}).  Thus it is not surprising that the frequencies (shown in Figs.~\ref{bmfig} and \ref{dmfig}) of both of these out-of-phase modes exhibit similar behavior as a function of temperature.

As discussed in Section~\ref{review}, the in-phase dipole mode considered in this Section is also an example of a locally isentropic mode.  Unlike the in-phase breathing mode, however, which is only isentropic at unitarity, the generalized Kohn mode is always characterized by $\bv_s = \bv_n$, at all temperatures and interaction strengths.  This feature also follows from the condition given in Eq.~(\ref{decouplecondition}) since $\bnab\cdot\bu = 0$ for the dipole mode.

\section{Concluding remarks}

In this paper, we have presented results for the breathing and dipole mode solutions of the Landau two-fluid equations for a Fermi superfluid at unitarity in an isotropic trap.  Our work is based on a recent variational formulation~\cite{TaylorPRA05} of the two-fluid equations.  We have shown that the variational equations simplify at unitarity, where the coefficients only depend on the compressibility and the superfluid density.  Understanding the nature of Fermi gases at unitarity, where the $s$-wave scattering length diverges, is a challenging many-body problem.  In contrast to the in-phase dipole and breathing modes, which we have shown to be independent of temperature, the out-of-phase modes are very dependent on temperature.   Measurement of the out-of-phase mode frequencies will provide a sensitive test of current microscopic theories of a Fermi gas at unitarity, including the predictions of ``universal thermodynamics"~\cite{Ho04}.  In particular, our mode frequencies in Figs.~\ref{bmfig} and \ref{dmfig} show that the results are very dependent on the temperature dependence of the superfluid density.  It would be very useful to have a more accurate \textit{ab-initio} calculation of $\rho_s$ (such as Ref.~\cite{Trivedi06}).  

In a companion paper~\cite{HuLiuPRL08}, we show that the frequencies of these hydrodynamic modes can be measured using two-photon Bragg spectroscopy, a standard tool used to study excitations in trapped, ultracold quantum gases~\cite{DavidsonRMP05}.  

We emphasize that the results presented in this paper for the frequencies of the low-lying out-of-phase breathing and dipole modes are based on the simplest possible variational ansatz for these modes.  Our variational results provide an \textit{upper bound} on the exact frequencies for these modes.  In future work, we will discuss results based on an improved variational ansatz.  

While we have concentrated on the two-fluid modes at unitarity in the present paper, our general variational formulation of the two-fluid equations can be applied anywhere in the BCS-BEC crossover, as long as the interactions are sufficiently strong to ensure collisionally hydrodynamic behavior.  It would be interesting to consider the two-fluid modes of a strongly-interacting Bose-condensate of dimer molecules, on the BEC side of unitarity.  

We can use the thermodynamic functions discussed in Section~\ref{NSRsec} to evaluate the temperature dependence of first and second sound velocities at unitarity in a uniform gas based on the NSR theory.  However, the physics of second sound in a uniform gas~\cite{Heiselberg05} is quite different from the out-of-phase modes in a trapped gas discussed in the present paper.  One finds that at unitarity, the BCS Fermi excitations make the dominant contribution to the second sound velocity but as one goes to lower temperature, the increasing gap $\Delta_0$ (see, for example, Ref.~\cite{TaylorPRA07}) freezes out this contribution relative to the undamped bosonic excitations.  The end result is that as $T\rightarrow 0$, the second sound velocity increases and approaches $c/\sqrt{3}$, where $c$ is the Bogoliubov phonon velocity.  We will give a more complete discussion of second sound in a uniform gas in another publication.

\begin{acknowledgments}
We thank Eugene Zaremba and John Thomas for helpful discussions.  We also acknowledge discussions with Yan He about Refs.~\cite{Levin07,Levin07b}.  E.T. and A.G. are supported by NSERC of Canada. H.H. and X.-J.L. are supported by the National Natural Science Foundation of China Grant No. NSFC-10774190, the National Fundamental Research Program of China Grant Nos. 2006CB921404 and 2006CB921306, and the Australian Research Center Council of Excellence.
\end{acknowledgments}

\appendix
\section{First sound in superfluid $^4$He as a locally isentropic mode}
\label{appendixA}

In part A of Section~\ref{bmu}, we showed that at unitarity the in-phase breathing mode is a locally isentropic mode ($\bnab\delta T =0$), where the local superfluid and normal fluid velocities are equal, $\bv_s(\br,t) = \bv_ n(\br,t)$.  
It is useful to compare this analysis with the case of superfluid $^4$He, where first sound also describes a locally isentropic mode.

The two-fluid modes in \textit{uniform} superfluid helium are to a very good approximation given by~\cite{Landau41, Khalatnikov} $\bv_s = \bv_n$ and $\rho_{s0}\bv_s + \rho_{n0}\bv_n = 0$, corresponding to first and second sound, respectively.  First sound describes a locally isentropic density oscillation ($\delta T = 0$), while second sound describes a pure temperature ($\delta\rho = 0$) oscillation~\cite{Khalatnikov}.  Note that in a uniform superfluid, the condition $\bnab \delta T = 0$ is equivalent to $\delta T =0$ since a uniform oscillation of the temperature is impossible.  

The existence of a locally isentropic first sound mode in uniform $^4$He is also accounted for by the condition we give in Eq.~(\ref{decouplecondition}).  In a uniform system, all equilibrium thermodynamic quantities are independent of position and the term in the second line Eq.~(\ref{decouplecondition}) that involves the gradient of $(\partial P/\partial s)_{\rho}$ vanishes (recall that in a trapped superfluid, it only vanishes at unitarity).  Since Eq.~(\ref{decouplecondition3}) is not satisifed by the plane-wave solutions of the uniform two-fluid equations, we see that the condition for a locally isentropic first sound mode to exist is given by Eq.~(\ref{decouplecondition2}), namely that $(\partial P/\partial s)_{\rho} = 0$.  Using the identity (see Sec.~16 in Landau and Lifshitz~\cite{LLSM})
\bea \left(\frac{\partial P}{\partial s}\right)_{\rho} = \frac{T}{\rho \bar{c}_v}\left(\frac{\partial P}{\partial T}\right)_{\!\rho}, \eea
where $\bar{c}_v = T(\partial\bar{s}/\partial T)_{\rho}$ is the equilibrium specific heat per unit mass, one sees that $(\partial P/\partial s)_{\rho} \simeq 0$ implies $(\partial P/\partial T)_{\rho} \simeq 0$.   In this case, the adiabatic and isothermal compressibilities are equal [$(\partial P/\partial\rho)_{\bar{s}} \simeq (\partial P/\partial\rho)_T$].  When dealing with the two-fluid equations in superfluid helium, this equivalence leads to a well known simplification in the equations for first and second sound~\cite{Khalatnikov}.  The simplified equations can be easily solved leading to the result that first sound is a locally isentropic mode ($\bv_s = \bv_n$), with a sound speed given by the adiabatic compressibility $u_1 = \sqrt{(\partial P/\partial\rho)_{\bar{s}}}$.

\section{Frequencies close to $T=0$ in the BCS approximation}
\label{appendixD}

In this Appendix, we show that the local density approximation is not responsible for the diverging frequency of the out-of-phase breathing and dipole modes as $T\rightarrow 0$ (see Figs.~\ref{bmfig} and \ref{dmfig}). In this limit, the reduced mass
moment is approximately the mass moment of the normal component, $M_r \rightarrow M_n$. Thus, the
out-of-phase mode frequency is inversely proportional to the normal mass
moment, which becomes very small at low temperature. One may question the numerical accuracy of the calculations. In particular, the strong temperature dependence of the out-of-phase mode frequencies at low $T$ might be an
artifact of the local density approximation used in Figs.~\ref{bmfig} and \ref{dmfig}. To check this point, we
calculate the out-of-phase breathing mode frequency using the thermodynamic functions for a trapped gas given by directly solving the Bogoliubov-de Gennes (BdG) equations for a finite number of atoms.  

We solve the coupled BdG
equations for the Bogoliubov quasiparticles of a Fermi gas in an isotropic
harmonic trap at unitarity.  A microscopic expression of the superfluid density
of a finite size inhomogeneous system may be derived by considering the
moment of inertia of the Fermi gas, or equivalently, by calculating the increase in free
energy after imposing a twisted boundary phase for the order parameter~\cite{TaylorPRA06}. In an isotropic trap, the superfluid density is given by
\begin{eqnarray}
\rho _{s0}\left( r\right)  &=&\rho _0\left( r\right) -\frac{\hbar ^2}{r^4}%
\sum_{nl}\left[ -\frac{\partial f\left( E_{nl}\right) }{\partial E_{nl}}%
\right] \times   \nonumber \\
&&\frac{l\left( l+1\right) \left( 2l+1\right) }{8\pi }\left[ u_{nl}^2\left(
r\right) +v_{nl}^2\left( r\right) \right] \label{nsiso}.
\end{eqnarray}
Here $u_{nl}\left( r\right) $ and $v_{nl}\left( r\right) $ are the radial
wavefunctions of the Bogoliubov quasiparticles.  The full wavefunctions
have the form $u_j\left( {\bf r}\right) =\left[ u_{nl}\left( r\right)
/r\right] Y_{lm}\left( \theta ,\phi \right) $ and $v_j\left( {\bf r}\right)
=\left[ v_{nl}\left( r\right) /r\right] Y_{lm}\left( \theta ,\phi \right) $.
The quasiparticle energy $E_{nl}$ is allowed to be negative, and the
summation of the level indices ($nl$) is over both positive and negative
energy levels. $f\left( x\right) $ is the Fermi-Dirac distribution
function.

\begin{figure}
\begin{center}
\epsfig{file=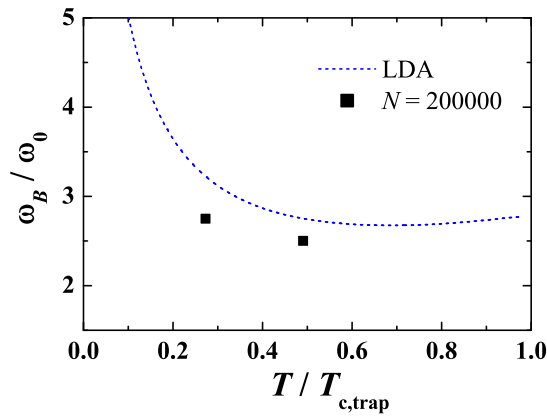, angle=0,width=0.40\textwidth}
\caption{Comparison of an LDA calculation of the frequency of the out-of-phase breathing mode with a calculation using the Bogoliubov-de Gennes equations. Here the LDA result
(solid line) is obtained by using the BCS mean-field theory for
a Fermi gas at unitarity.  The solid squares are calculated using the self-consistent Bogoliubov-de Gennes
mean-field equations.}
\label{bdgbmfig}
\end{center}
\end{figure}

Fig.~\ref{bdgbmfig} compares the frequency of the out-of-phase breathing mode calculated (a) within LDA using a mean-field BCS equation of state and (b) from a self-consistent calculation of the BdG equations.  The BdG frequency (for $N=2\times 10^5$ atoms) is smaller but close to the LDA result.  The increase of the mode frequency with decreasing temperature is clearly seen in both the BdG and LDA calculations. The disagreement seen in Fig.~\ref{bdgbmfig} is not unexpected since numerical calculations we carried out as a function of $N$ show that the BdG results converge very slowly with increasing $N$.  We 
conclude that the strong increase of the out-of-phase mode frequency in the
low temperature regime is not an artifact of the LDA.

\section{Frequencies close to $T_c$ in the BCS approximation}
\label{appendixC}

Close to the superfluid transition temperature, the temperature dependence
of the out-of-phase mode frequencies using the mean-field BCS superfluid density can be worked out without using the LDA.  In this
region, an analytic result for the weak-coupling BCS superfluid density in a trapped gas is given by Baranov and Petrov~\cite{Baranov}.
Assuming a second order phase transition near $T_c$, Ginzburg-Landau theory predicts the following temperature dependence for the position dependent order parameter in
a harmonic trap~\cite{Baranov}: 
\begin{equation}
\Delta (r)\propto T_c\left( \frac{T_c-T}{T_c}\right) ^{1/2}\left( 1-\frac{r^2%
}{R_c^2}\right) ^{1/2}.
\end{equation}
Here the radius $R_c\propto \sqrt{\delta T/T_c}$ and $\delta T\equiv T_c-T$. Since 
the BCS superfluid density varies as $\rho_{s0}\left( r\right) \propto \Delta ^2(r)$ near 
$T_c$, we may write 
\begin{equation}
\rho _{s0}\left( r\right) =\alpha \left( 1-\frac{r^2}{R_c^2}\right) ,
\label{nsclosetotc}
\end{equation}
where the prefactor $\alpha \propto \delta T/T_c$. We note that even though
both $\alpha $ and $R_c^2$ vanish linearly with temperature close to $T_c$, the ratio $\alpha /R_c^2$\ remains finite.

In the vicinity of $T_c$, the superfluid mass moment is much smaller than
the normal mass moment, and the reduced mass moment of both the dipole and breathing modes reduces to the superfluid mass moment, $M_r \rightarrow M_s$%
. Taking the breathing mode as an example, its mode frequency [given by Eq.~(\ref{mode2iso})] reduces to $\omega _{2B}^2=k_s^B/M_s^B$ (note that $k^B_s/M^B_s$ remains finite as $T\rightarrow T_c$, while $M^B_s/M^B_n$ vanishes).  Since $R_c\ll 1$, the adiabatic
compressibility is nearly constant in the region of interest, and we
denote it as $\gamma =(\partial \mu /\partial \rho )_s$ for $T\rightarrow
T_c$.

The calculations of the superfluid mass moment and the spring constant are
straightforward. Substituting Eq. (\ref{nsclosetotc}) into Eqs. (\ref{Mstildeiso}%
) and (\ref{ksiso}), we obtain
\begin{equation}
M_s^B=\frac{8\pi }{35}\alpha R_c^5
\end{equation}
and 
\begin{equation}
k_s^B=\frac{16\pi }7\gamma \alpha ^2R_c^3.
\end{equation}
Thus, using a weak-coupling BCS mean-field calculation near $T_c$, the frequency of the out-of-phase breathing mode is predicted to be
\begin{equation}
\omega _{2B}^2=\frac{k_s^B}{M_s^B}=10\gamma \left( \frac \alpha {R_c^2}%
\right) .  \label{w2btc}
\end{equation}
As noted earlier, both $\alpha /R_c^2$ and $\gamma $ approach constant values close to $T_c$. Thus, the out-of-phase breathing mode frequency
is finite at the transition temperature. We have checked the validity of Eq.
(\ref{w2btc}) by numerically calculating the values of $\alpha $, $\gamma $,
and $R_c$. As these parameters do not change much above the temperature $%
0.5T_{c,trap}$, the mode frequency becomes fairly constant in this temperature range, in agreement with our LDA results in Fig.~\ref{bmfig} for the breathing-mode frequency based on the scaled BCS superfluid density (dashed curve).


\begin{thebibliography}{99}
\bibitem{note2} In addition to the assumption that two-fluid hydrodynamics is applicable, we have also made use of the scaling ansatz in Eq.~(\ref{bmansatz}) in arriving at the result in Eq.~(\ref{mode1iso}).  As shown in Ref.~\cite{Castin}, there is an exact eigenstate of the isotropic trap Hamiltonian such that all atoms move with velocity $\bv(\br,t) = a\br\cos \omega t$.
\end{thebibliography}
\end{document}